\documentstyle[sprocl]{article}

\input{psfig}
\def\be{\begin{equation}}
\def\ee{\end{equation}}
\def\bea{\begin{eqnarray}}
\def\eea{\end{eqnarray}}
\begin{document}

\title{ROLE OF DEFECTS AND IMPURITIES IN DOPING OF GaN}

\author{J\"org Neugebauer\footnote{present address: Fritz-Haber-Institut der
    MPG, Faradayweg 4--6, D-14195 Berlin, Germany} and Chris~G.~Van de Walle}

\address{ Xerox Palo Alto Research Center, 3333 Coyote Hill Road, Palo Alto,
  CA 94304, USA}


\maketitle 
\abstracts{ We have calculated formation energies and position of
  the defect levels for all native defects and for a variety of donor and
  acceptor impurities employing first-principles total-energy calculations.
  An analysis of the numerical results gives direct insight into defect
  concentrations and impurity solubility with respect to growth parameters
  (temperature, chemical potentials) and into the mechanisms limiting the
  doping levels in GaN.  We show how compensation and passivation by native
  defects or impurities, solubility issues, and incorporation of dopants on
  other sites influence the acceptor doping levels.}

\section{Introduction}
GaN has attracted widespread attention for producing highly-efficient blue
light-emitting diodes~\cite{nakamura94} and as a promising candidate for
high-temperature or high-power devices. Recent progress in growth and device
techniques enabled the fabrication of first prototypes of blue light-emitting
laser-diodes. However, despite the rapid progress in the development of
GaN-based devices doping is still an important issue. Particularly the lack of
high $p$-type doping levels limits the achievable injection currents important
for improving e.g. laser diodes. One reason why high $p$-type doping levels
are difficult to achieve is the high ionization energy characteristic for
acceptors in GaN. Magnesium which is the most commonly used acceptor in GaN
has an acceptor binding energy of about 160\,meV~\cite{akasaki91} implying that
at room temperature the hole concentration is less than 1\% of the Mg
concentration.

The hole concentration is given by \mbox{$n_{\rm hole}=N_{\rm acceptors}
  \exp{(-E_a/k_b T)}$} where $N_{\rm acceptors}$ is the number of acceptors,
$E_a$ the acceptor ionization energy, $k_b$ the Boltzman constant, and $T$ the
device temperature. The hole
concentration can therefore be increased by: (i) increasing the accceptor
concentration, (ii) increasing the device temperature, and (iii) finding new
acceptors with a lower ionization energy. Increasing the temperature is a
device dependent issue and will not be discussed here. One
should however keep in mind that increasing the temperature above room
temperature commonly {\em decreases} the carrier mobility. 

Decreasing the acceptor ionization energy requires to find new acceptor
species. There is some experimental evidence that acceptor levels with a lower
binding energy exist: possible candidates which are discussed in the
literature are C~\cite{fischer95}, Ca~\cite{lee96}, and Zn. However,
for those elements to be really useful for doping it is essential that
sufficiently high acceptor concentrations can be achieved. 

In the present paper we will discuss which mechanisms limit the doping levels
in GaN. In particular compensation by native defects and impurities, complex
formation, solubility issues, and incorporation on electrically not active
sites will be studied. Based on these results we explain, why growing in
hydrogen-rich conditions is beneficial on acceptor incorporation and how this
mechanism can be used to optimize $p$-type doping in GaN. As possible acceptor
species Mg (which is the most commonly used acceptor), Zn, and Ca will be
investigated.

\section{Method}

The equilibrium concentration $c$ of an impurity or defect at temperature
$T$ is determined by its formation energy, $E^f$:
\begin{equation}\label{eq:def_conc}
c = N_{\rm sites} \,\, \exp^{S/k_B} \exp^{-E^f/k_B\,T}
\end{equation}
where $N_{\rm sites}$ is the number of sites the defect can be built
in. 
$k_B$ is the Boltzmann constant and $S$ the vibrational entropy. The
vibrational entropy is, at the present stage of our work, not
explicitly included, which would be computationally very demanding.
Entropy contributions cancel to some extent~\cite{qian92}, and are
small enough not to affect any qualitative conclusions.  The effect of
entropy can be included in an approximate way; experimental and
theoretical results show that the entropy $S$ is typically in the
range between 0 (no entropy contributions) and 10\,$k_B$ (where $k_B$
is the Boltzman constant).\cite{laks92}

The formation energy depends on various parameters. For example, the
formation energy of a Mg acceptor is determined by the relative
abundance of Mg, Ga, and N atoms. In a thermodynamic context these
abundances are described by the chemical potentials $\mu_{\rm Mg}$,
$\mu_{\rm Ga}$, and $\mu_{\rm N}$. If the Mg acceptor is charged, the
formation energy depends further on the Fermi level ($E_F$), which
acts as a reservoir for electrons. Forming a substitutional Mg
acceptor requires the removal of one Ga atom and the addition of one
Mg atom; the formation energy is therefore:
\begin{equation}\label{eq:mf_form}
E^f({\rm \mbox{GaN:Mg}}_{\rm Ga}^q) = E_{\rm tot}({\rm \mbox{GaN:Mg}}_{\rm Ga}^q) 
    - \mu_{\rm Mg} + \mu_{\rm Ga} + q E_F
\end{equation}
where $E_{\rm tot}({\rm \mbox{GaN:Mg}}_{\rm Ga}^q)$ is the total
energy derived from a calculation for substitutional Mg, and $q$ is
the charge state of the Mg acceptor.  Similar expressions apply to the
hydrogen impurity, and to the various native defects. For calculating
the total energies we have performed first-principles calculations
based on density-functional theory (DFT) using a supercell approach
with 32 atoms per cell, a plane-wave basis set with 60\,Ry cutoff and
soft Troullier-Martins pseudopotentials.\cite{troullier91} Details of
the method and convergency checks can be found
elsewhere.\cite{prb1_94,stumpf94,mrs95_tb}\\[1ex]

\section{Compensating centers}
\subsection{Native defects\label{sec:nat_def}}

We have calculated the position of the defect levels and the formation
energies for all native defects in GaN: vacancies ($V_{\rm Ga}$, $V_{\rm N}$),
antisites (Ga$_{\rm N}$, N$_{\rm Ga}$), and interstitials (Ga$_i$, N$_i$). All
relevant charge states were taken into account. From these results we can
classify the defects into donors, acceptors and amphoteric defects: V$_{\rm
  N}$ and Ga$_i$ are donors, V$_{\rm Ga}$ is an acceptor, and N$_i$ and both
antisites are amphoteric.

We find that antisites and interstitials have high formation energies, and are
therefore very unlikely to occur in any significant
concentration.\cite{prb1_94} The vacancies, however, have lower formation
energies. Under conditions of thermodynamic equilibrium, low formation
energies are required for the defect to occur in high concentration. In
particular, the nitrogen vacancy (a single donor) has a low formation energy
under $p$-type and semiinsulating conditions, and the Ga vacancy (a triple
acceptor) gets a low formation energy under $n$-type conditions. Thus, based
on the assumption of thermodynamic equilibrium, the dominant native defects
are the vacancies: in $p$-type GaN the N vacancy, in $n$-type GaN the Ga
vacancy.

The fact that the N vacancy (a donor) has a low formation energy under
$p$-type conditions implies that it can be an efficient compensating center
for acceptors. Similarily, under $n$-type conditions the Ga vacancy (an
acceptor) has a low formation energy and can compensate donors. There is also
strong evidence that the Ga vacancy or donor-Ga-vacancy complexes are involved
in the characteristic yellow luminescence observed in the photo-luminescence
of GaN.\cite{apl96_yl}

\subsection{Hydrogen}

Hydrogen is highly abundant in many of the high-temperature growth techniques
such as MOCVD or HVPE (hydride vapor phase epitaxy). There is also strong
experimental evidence that H is involved in the $p$-type doping of
GaN.\cite{nakamura92}

Let us briefly recall the properties of monatomic and molecular H in
GaN.\cite{prl95_hy,apl95_hy} H exhibits a fundamental difference in the
behavior between $p$-type and $n$-type material. In $p$-type material H
occupies the N-antibonding site, it behaves as a donor (it is positively
charged), has a high solubility and a high diffusivity. In $n$-type conditions
H is negatively charged (it is an acceptor) and a Ga-antibonding site is
prefered. Compared to $p$-type conditions the diffusivity and solubility are
dramatically reduced. These properties render H as an important compensating
center: under $p$-type conditions it can efficiently compensate acceptors,
under $n$-type conditions it can (less efficiently due to the lower
solubility) compensate donors.

Our calculations show further that neutral H is unstable against forming H$^+$
or H$^-$ for all Fermi-level positions.  This behavior is characteristic for a
negative $U$ center. A negative-$U$ behavior was also found for H in
Si~\cite{vandewal89} and GaAs~\cite{pavesi92}; however, the value calculated
for GaN ($U=-2.4$\,eV)~\cite{prl95_hy} is unusually large, and, to our
knowledge, larger than any measured or predicted value for any defect or
impurity in any semiconductor.

We have also investigated hydrogen molecules in GaN.\cite{prl95_hy} The
results show that H$_2$ is unstable with respect to dissociation into
monatomic hydrogen. The formation energy is much higher than that of
H$_2$ in vacuum. Both features, the low stability of the hydrogen
molecule and its unfavorably high formation energy, are distinct
properties of GaN and very different from the case of Si or GaAs.

\section{$p$-type doping in GaN}
\subsection{Solubility}

The solubility of dopants in GaN is limitted by the formation of dopant
clusters or the formation of chemical compounds between the acceptor species
and N or Ga atoms. Our calculations show that the solubility of Mg is limitted
by the formation of Mg$_3$N$_2$; the reason is the high chemical stability of
this compound (formation enthalpy: $\Delta H_f$=4.8\,eV). For Ca the
solubility is limitted by Ca$_3$N$_2$ ($\Delta H_f$=4.6\,eV). For Zn acceptors
the solubility is limitted by two phases: under Ga-rich conditions by Zn-bulk,
under N-rich conditions by Zn$_3$N$_2$. The origin is the very low formation
enthalpy of Zn$_3$N$_2$ ($\Delta H_f$=0.2\,eV) which implies that this phase
can be formed only under N-rich conditions.

For the following discussion we will fix the chemical potentials. For the Ga
chemical potential we assume Ga-rich conditions (which appear to be common in
experimental growth conditions). For the acceptors we set the chemical
potential to its upper limit (we are interested in the maximum achievable
acceptor concentration) which is given by their solubility limit. Using the
same arguments we set the H chemical potential to that of H$_2$ molecules.
Using these values for the chemical potentials the formation energy is solely
a function of the Fermi energy.

\subsection{Compensation}

\setlength{\unitlength}{1mm}
\begin{figure}[tb]
  \begin{picture}(80,40)(-3,10)
    \psfig{figure=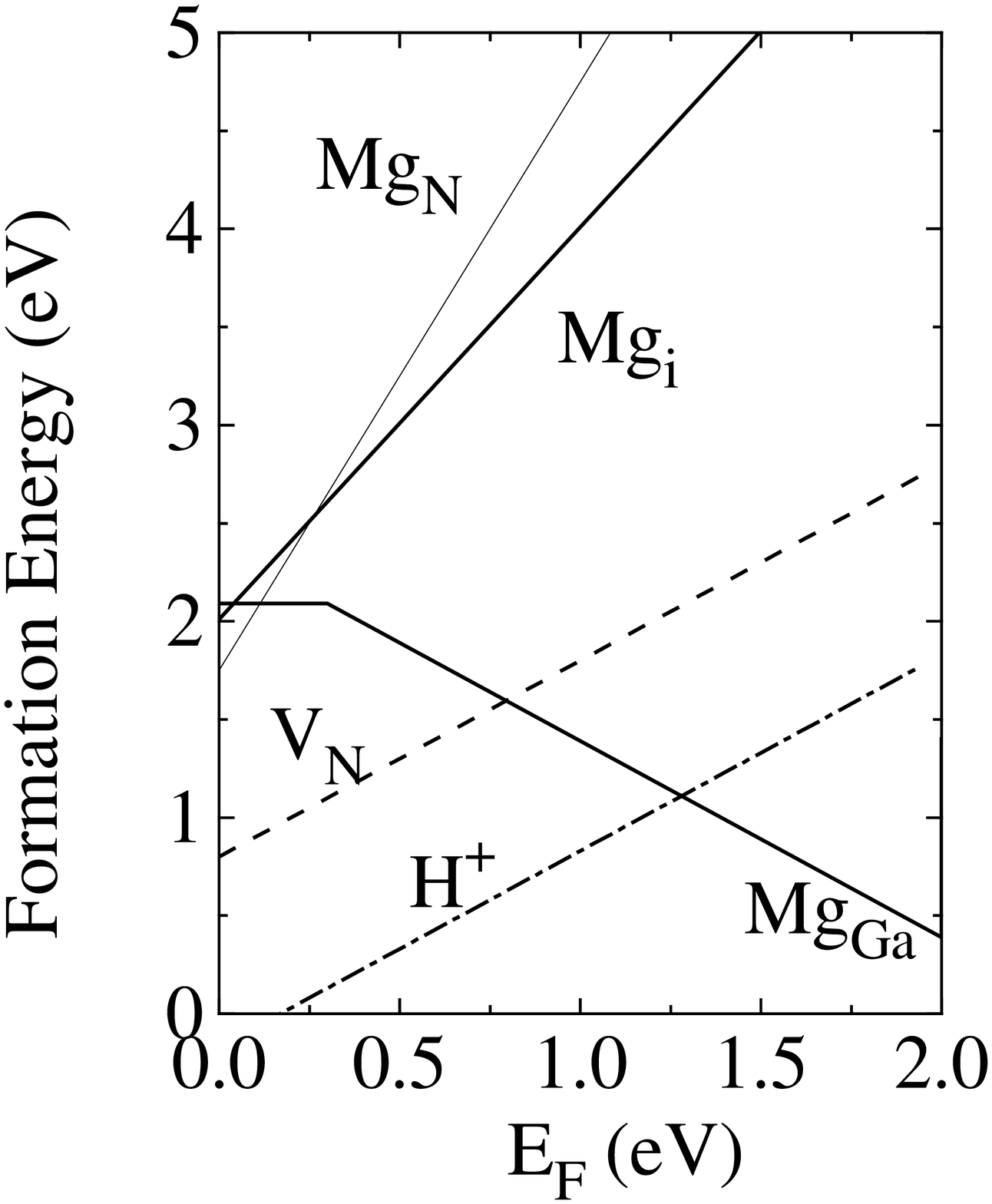,height=5.5cm}
    \psfig{figure=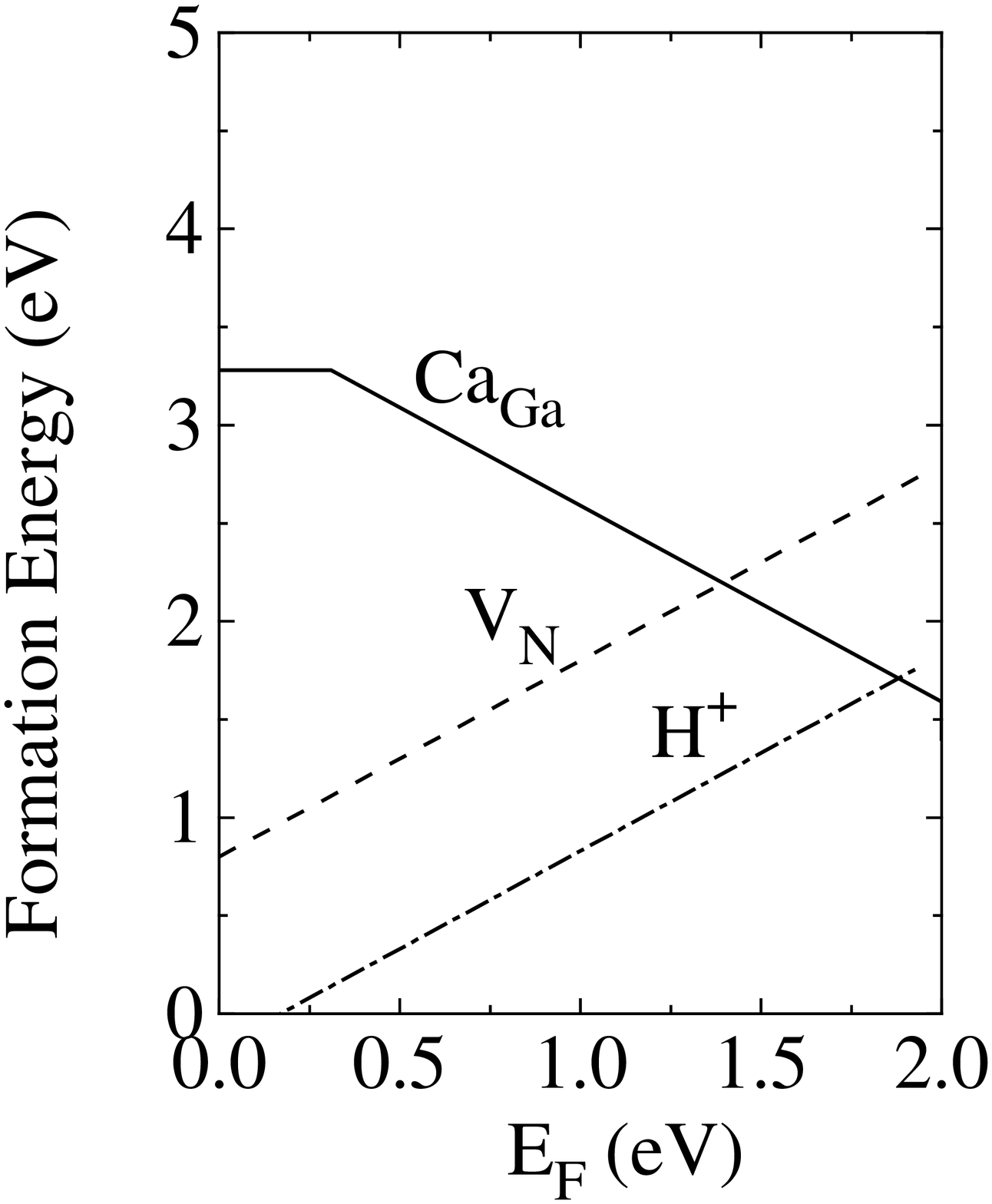,height=5.5cm}
    \psfig{figure=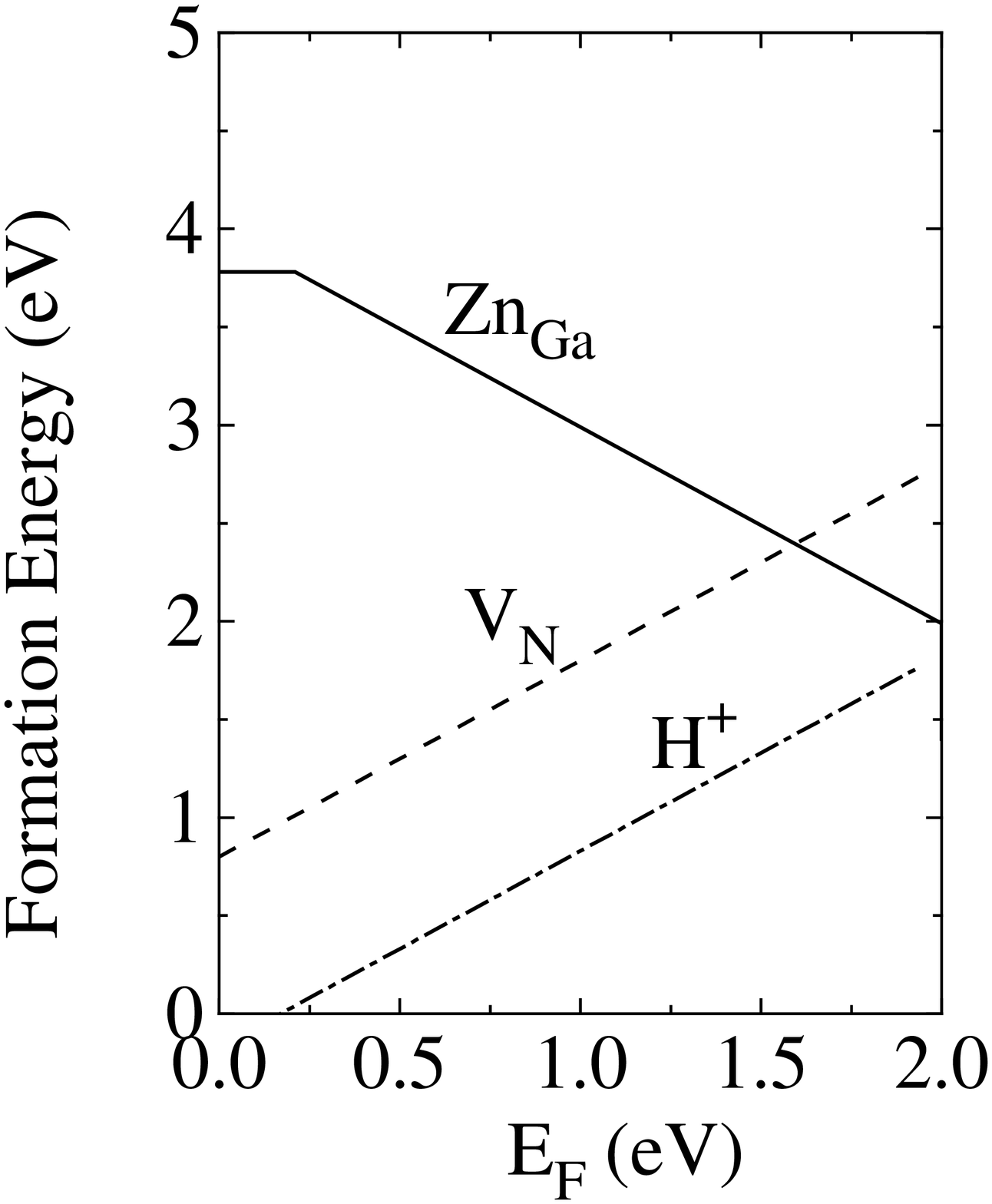,height=5.5cm}
  \end{picture}
\caption{Formation energy as a function of the Fermi level for (a) Mg, (b) Ca,
  and (c) Zn in different configurations (Ga-substitutional, N-substitutional,
  and interstitial configuration). Also included are the native defects and
  interstitial H.
\label{fig:form_erg}}
\end{figure}

Before discussing the role of H in $p$-type doping let us briefly focus on the
H free case, i.e., only the acceptor and native defects are present.
Figure~\ref{fig:form_erg} shows the corresponding formation energies as
function of the Fermi energy. As pointed out in Sec.~\ref{sec:nat_def} the
dominant native defect under $p$-type conditions is the N vacancy; all other
defects are higher in energy and outside the plotrange. The slope of the
formation energies characterizes the charge state; a positive slope (as e.g.
found for the N vacancy) indicates a positive charge state, corresponding to a
donor.

Using the calculated formation energies, and taking into account that the
Fermi energy is fixed by the condition of charge neutrality, the equilibrium
concentration [Eq.~(1)] for each defect can be calculated as function of
temperature. At temperatures exceeding 1000\,K the N vacancies increasingly
compensate the acceptors.\cite{apl95_hy} Low temperature growth techniques
such as MBE should suffer less from this problem. This conclusion would be
consistent with the fact that only in MBE-grown GaN $p$-type conductivity can
be achieved without any post-growth processing.\cite{molnar93,lin93}

\subsection{Role of H in achieving $p$-type doping}

H-rich conditions are characteristic for many of the high-temperature growth
techniques such as MOCVD and HVPE. Figure~\ref{fig:form_erg} shows that under
these conditions H becomes the dominant donor; the formation energy of H is in
the considered interval of Fermi energies always lower than the dominant
native defect, the N vacancy. From our calculated equilibrium concentrations
we find that the acceptor and H concentration are for all temperatures
virtually identical indicating that H is a very efficient compensating center
which completely compensates the acceptors. Further, compared to the H-free
case the concentration of acceptors is {\em increased} and the defect
concentration is {\em decreased}. Both effects are observed for all three
acceptors (Mg, Ca, Zn) and are crucial to increase doping levels.

What is the mechanism by which H changes the acceptor and defect
concentration? In a plot such as Fig.~\ref{fig:form_erg}, the Fermi level
position can be roughly estimated to be near the crossing point between the
acceptor and the dominant donor species. At this point their formation
energies (and hence there concentrations) are equal, ensuring charge
neutrality.\cite{neutral} By going from H-free conditions to H-rich conditions
the crossing point shifts to higher Fermi energies. An increase in the Fermi
energy {\em generally} decreases the formation energy of acceptors and
increases the formation energy of donors (defects), thus resulting in a
lowered defect concentration and an increased acceptor concentration. We note,
that this mechanism works only if H is able to significantly shift the Fermi
energy which is the case if (i) H is the dominant donor (i.e. its formation
energy must be lower than that of all native defects) and (ii) its formation
energy must be comparable to that of the dopant impurity (a crossing point
must exist in the band gap) which is the case for all three acceptors
considered here.

Despite the fact that growing under H-rich conditions improves acceptor and
defect concentrations the acceptors are almost completely compensated by the H
impurities. Therefore, growing under H-rich conditions results in
semiinsulating material consistent with experimental
observations.\cite{nakamura92} In order to activate the acceptors, post-growth
treatments are necessary to eliminate the compensation by H. 

For Mg dopants the H donors and Mg acceptors can actually form electrically
neutral complexes with a binding energy of $\approx 0.7$\,eV.\cite{prl95_hy}
Since the complex is mainly characterized by a strong N-H bond~\cite{prl95_hy}
we expexct similar complexes also for Zn and Ca acceptors. For the specific
choice of chemical potentials made here, this binding energy is low enough for
the complexes to be dissociated at the growth temperature; however, the Mg and
H will form pairs when the sample is cooled to room temperature, consistent
with experimental observations.\cite{goetz95}

The first step in the activation process is the dissociation of the H-acceptor
complex. Our estimated dissociation barrier for the Mg-H complex is
1.5\,eV, calculated by considering a jump to a nearest-neighbor site;
the total barrier may be slightly higher.\cite{prl95_hy} This barrier
should be low enough to be overcome at modest annealing temperatures
(around 300$^{\rm o}$C).
Experimental results show, however, that activation
has to be carried out at much higher temperatures ($>600^{\rm
o}$C).\cite{nakamura92} The reason is that dissociation alone is
insufficient; in order to prevent the H from compensating the Mg
acceptor it has to be either removed (to the surface or into the
substrate) or neutralized (e.g., at an extended defect).

The calculated diffusion barrier for H$^+$ in GaN is low ($\approx
0.7$\,eV~\cite{prl95_hy}) indicating that H$^+$ is highly mobile and can easily
migrate to the surface or extended defects. The high temperature necessary to
activate the Mg and Ca acceptors therefore reflects an activation barrier for
eliminating H as compensating center by incorporating it at extended defects
(which typically occur in high concentrations in GaN~\cite{lester94}) or by
removal of H through desorption at surfaces.

\subsection{Incorporation on other sites}

Another mechanism, that may limit the hole concentration, is self-compensation
of the acceptors: instead of being incorporated on the Ga substitutional site
the acceptor may be built in on other sites where it is electrically inactive
or even becomes a donor. As possible configurations we have investigated the N
substitutional site and several interstitial configurations. For all three
acceptors considered here the behavior is very similar. We will therefore
focus our discussion on Mg. 

The calculated formation energies are displayed in Fig.~\ref{fig:form_erg}.
The positive slope in the formation energy indicates that Mg in both
configurations acts as a donor: Mg$_i$ as a double donor, Mg$_{\rm N}$ as a
triple donor. Whereas the behavior of Mg$_i$ is expected, the donor character
for Mg$_{\rm N}$ is not as obvious: at first sight Mg should act as a triple
acceptor. The replacement of the ``small'' N atom ($r_{\rm cov}=0.75$\,\AA)
with a ``large'' Mg atom ($r_{\rm cov}=1.36$\,\AA) results in a large increase
of the nearest neighbor bond length (by 24\% !), giving rise to substantial
changes in the positions of the defect levels. The large strain around the Mg
atom further explains the high formation energy, rendering this site
energetically unfavorable. We therefore conclude that Mg (and also Zn and Ca)
will always prefer the Ga substitutional site: incorporation on other sites
can be ruled out. For other possible acceptors (particularly elements with a
small ionic radius), the situation may be different.

\section{Conclusions}

Based on first-principles calculations we have studied several mechanisms
which may limit the hole concentration in GaN doped with Mg, Ca, and Zn.  Two
mechanisms are found to be important: (i) solubility issues (the formation of
Mg$_3$N$_2$, Ca$_3$N$_2$, and Zn bulk) and (ii) the compensation by native
defects. Incorporation of Mg, Zn, and Ca acceptors on the N site or in an
interstitial configuration was found to be negligible. Combining our numerical
results about native defects, interstitial H and acceptor impurities we could
identify why H incorporation increases the acceptor concentration and
simultaneously reduces compensation by native defects.

\section*{References}
\bibliography{dft,gan,my,defects,footnotes,surface}
\bibliographystyle{mrssty}

\end{document}